\documentclass[aps,prd,preprintnumbers,showpacs,showkeys,nofootinbib,superscriptaddress,fleqn,floatfix,tightenlines,10pt]{revtex4-1}
\usepackage{amsmath,amsfonts,amssymb,amscd,amsxtra,amsthm}
\usepackage{graphicx}  % Include figure files
\usepackage{epstopdf}
\usepackage{dcolumn}  % Align table columns on decimal point
\usepackage{bm}          % bold math
\usepackage{slashed}
\usepackage[utf8]{inputenc}
\usepackage{CJK}
\usepackage{cancel}
\usepackage{mathtools}
\usepackage{amsbsy}
\usepackage{amstext}
\usepackage[normalem]{ulem} % \sout{old text} for strikeout
\usepackage[dvipsnames]{xcolor} % For blue in-text comments and
\usepackage{array}
\usepackage{slashed}
\renewcommand\sout{\bgroup \color{red} \ULdepth=-.5ex \ULset}

\usepackage[caption = false]{subfig}
\usepackage{multirow}
\begin{document}
\preprint{INHA-NTG-08/2018}
\title{Mass spectra of heavy mesons with instanton effects}

\author{Qian Wu}
\email[E-mail: ]{wuqian@smail.nju.edu.cn}
\affiliation{Department of Physics, Nanjing University, Nanjing
  210093, China}

\author{Emiko Hiyama}
\email[E-mail: ]{hiyama@riken.jp}
\affiliation{RIKEN Nishina Center,RIKEN,2-1 Hirosawa,351-0115
  Saitama,Japan}
\affiliation{Department of Physics,Kyushu
  University,819-0395,Fukuoka,Japan}

\author{Hyun-Chul Kim}
\email[E-mail: ]{hchkim@inha.ac.kr}
\affiliation{Department of Physics, Inha University, Incheon 22212,
  Republic of Korea}
\affiliation{Advanced Science Research Center, Japan Atomic Energy
  Agency, Shirakata, Tokai, Ibaraki, 319-1195, Japan} 
\affiliation{School of Physics, Korea Institute for Advanced Study
  (KIAS), Seoul 02455, Republic of Korea}

\author{Ulugbek Yakhshiev}
\email[E-mail: ]{yakhshiev@inha.ac.kr}
\affiliation{Department of Physics, Inha University, Incheon 22212,
  Republic of Korea}

\author{Hongshi Zong}
\email[E-mail: ]{zonghs@nju.edu.cn}
\affiliation{Department of Physics, Nanjing University, Nanjing
  210093, China}
\affiliation{Joint Center for Particle, Nuclear Physics and Cosmology,
  Nanjing 210093, China}
\affiliation{State Key Laboratory of Theoretical Physics, Institute of
  Theoretical Physics, CAS, Beijing, 100190, China}

\begin{abstract}
  We investigate the mass spectra of ordinary heavy mesons, based on
  a nonrelativistic potential approach. The heavy-light quark
  potential contains the Coulomb-type potential arising from one-gluon
  exchange, the confining potential, and the instanton-induced
  nonperturbative local heavy-light quark potential. All parameters
  are theoretically constrained and fixed. We carefully examine the
  effects from the instanton vacuum. Within the present form of the
  local potential from the instanton vacuum, we conclude that the
  instanton effects are rather marginal on the charmed mesons. 
\end{abstract}
%\paces{12.38.Mh,12.39.-x,25.75.Nq}

\maketitle
\section{Introduction}
The structure of hadrons containing a heavy quark is systematically
understood when the mass of the heavy quark is taken to infinity. This
is valid, since the heavy-quark mass $m_Q$ is
much larger than the $\Lambda_{\mathrm{QCD}}$, i.e. $m_Q\gg
\Lambda_{\mathrm{QCD}}$. Then a new type of symmetry arises:
the physics is not changed by the exchange of the heavy-quark
flavor. This is called heavy-quark flavor symmetry. In this limit,
the spin of the heavy quark $\bm{S}_Q$ is conserved, which brings
about the spin conservation of the light degrees of freedom
$\bm{S}_L$. So, the spin of a heavy hadron is also conserved in this
limit: $\bm{S}=\bm{S}_L+\bm{S}_Q$. This is often called heavy-quark
spin symmetry~\cite{Isgur:1989vq, Isgur:1991wq, Georgi:1990um}. The
heavy quark is entirely decoupled from the internal dynamics of a
heavy hadron in the limit of $m_Q\to\infty$ and the interaction among
light degrees of freedom becomes spin-independent. The infinitely
heavy-quark mass limit allows one to use the inverse of the
heavy-quark mass, $1/m_Q$, as an expansion parameter. The
spin-dependent part of the interaction appears as the next-to-leading
order in the $1/m_Q$ expansion, which is proportional
to $1/m_Q$ and stems from the chromomagnetic moment of the
quark (see, for example, reviews~\cite{Neubert:1993mb,
  Casalbuoni:1996pg, Bigi:1997fj, Chen:2016spr} and
books~\cite{Manohar:2000dt, Mannel:2004}).

In the limit of $m_Q\to\infty$, the classification of conventional
heavy meson states $Q\bar{q}$ with a single heavy quark $Q$ is rather
simple, where $\bar{q}$ denotes the light anti-quark constituting the heavy
meson. Since the heavy quark is decoupled in the $m_Q\to \infty$ limit,
the flavor structure is solely governed by the light quarks. Thus the
lowest-lying states of the heavy meson is classified as the
antitriplet meson $\overline{\bm{3}}$. Moreover, the mesons with spin
$s=0$ and those with $s=1$ are found to be degenerate, so that the
pseudoscalar and vector heavy mesons consist of the doublets in the
limit of $m_Q\to\infty$. This degeneracy is lifted by introducing the
spin-dependent interactions coming from $1/m_Q$ order. Based on this
heavy-quark flavor-spin symmetry, there has been a great deal of
theoretical works on properties of both the lowest-lying and excited
heavy mesons: lattice QCD~\cite{McNeile:2012qf, Moir:2013ub,
  Kalinowski:2015bwa, Cichy:2016bci, Cheung:2016bym, Guo:2018kno}, the
nonrelativistic and relativistic quark
models~\cite{Mukherjee:1993hb, Godfrey:2015dva, Hiorth:2002pp,
  Guo:2012tm, Eshraim:2014eka}, potential
models~\cite{Zeng:1994vj, Matsuki:1997da, Lahde:1999ih, Barnes:2005pb,
  Vijande:2006hj, Radford:2009bs}, QCD sum rules~\cite{Bagan, Ball,
  Dai} , holographic QCD~\cite{Branz:2010ub}, and so on.

The potential models for the heavy mesons are usually based on two
important physics: the quark confinement and the perturbative
one-gluon exchange. While these two ingredients of the potentials
describe successfully both properties of quarkonia and heavy mesons,
certain nonperturbative effects need to be considered. Diakonov et
al. derived the central part of the heavy-quark potential from the
instanton vacuum, using  the Wilson loop~\cite{Diakonov:1989un}. The
spin-dependent part can be easily constructed by employing the
Eichten-Feinberg formalism~\cite{Eichten:1980mw}. The effects of the
heavy-quark potential from the instanton were examined only
very recently by computing the quarkonium
spectra~\cite{Turimov:2016adx}.  
The results showed that the effects of the instanton turn out to be
rather small on the quarkonium spectra. Chernyshev et al. investigated
the effects of a random gas of instantons and anti-instantons on
mesons and baryons containing one or several heavy
quarks~\cite{Chernyshev:1995gj}. They first derived the \emph{local}
effective interactions from the random instanton-gas model (RIGM) and
then employed them to estimate the heavy-hadron mass spectra within a
simple variational method, including the harmonic oscillator
potential as a simple expression of the quark confinement . They
obtained results in qualitative agreement with the experimental data
on the low-lying heavy mesons. However, it is of great importance to
examine cautiously such nonperturbative effects on the heavy hadron
spectra in a quantitative manner.  

In the present work, we aim at exploring carefully the heavy-light
quark potentials, which were derived from the RIGM, examining their
effects on the mass spectra of the heavy mesons. For simplicity and
convenience, we will use the nonrelativistic framework in dealing with
the heavy-light quark interactions from the RIGM. In any potential
models for describing the quarkonia and heavy mesons, there are two
essential components: the quark confinement and the one-gluon exchange
contribution, which we want to introduce in addition to the interaction 
from the instantons. Instead of a simple variational method used
in Ref.~\cite{Chernyshev:1995gj}, we employ a more elaborated and
sophisticated framework, i.e. the Gaussian expansion
method (GEM), which is well known
for the successful description of 
two- and few-body systems~\cite{Kamimura1988, Hiyama:2003cu,
  Hiyama2004, Hiyama2009}, so that we reduce numerical uncertainties
arising from the simple variational method. 
As will be shown in this work, the present form of the heavy-light
quark interaction based on the RIGM has only marginal effects on the
mass spectra of the heavy mesons.  The quark potentials of one-gluon 
exchange and the quark confinement already reproduce 
approximately the experimental data on the spectra of the low-lying
heavy mesons. However, since the heavy-mesons contain a light quark,
we still expect that certain nonperturbative effects will come into
play. We will discuss them also in the present work.  

This paper is organized as follows: In Section II, we define the
heavy-light quark potentials arising from one-gluon exchange and the
quark confinement. We then introduce the effective potential coming
from the nonerpturbative heavy-light quark interactions based on the
RIGM. In Section III, we show how to solve the nonrelativistic
Schr\"odinger equation with the heavy-light quark potential within the
framework of the GEM. That 
will be the framework for numerical calculations in the present work.
In Section IV, we present the results and discuss them in comparison
with the experimental data. The final Section is devoted to summary
and conclusion. We also discuss a possible future outlook. 

\section{Heavy-light quark potential}
The general structure of the heavy-light quark potentials is expressed
as
\begin{align}
  \label{eq:1}
V(r)= V_c (r)+V_{SS}(r)(\bm{S}_Q\cdot
\bm{S}_q)+V_{LS}(r)(\bm{L}\cdot\bm{S}) + V_{T}(r) [3(\bm{S}_Q\cdot
\hat{\bm{n}})(\bm{S}_q\cdot \hat{\bm{n}})- \bm{S}_1\cdot \bm{S}_2],
\end{align}
where $V_c$ is the central part of the potential. The $V_{SS}$,
$V_{LS}$, and $V_T$ are called respectively the spin-spin term, the
$LS$ term that shows the coupling between the orbital angular momentum
and the spin angular momentum, and the tensor term.
Following Ref.~\cite{Eichten:1980mw}, the spin-dependent potential is
derived from the central potential. $\bm{S}_Q$ and $\bm{S}_q$ denote
the spin operators for the heavy and light quarks,
respectively. $\bm{L}$ and $\bm{S}$ represent respectively the
operator of the relative orbital angular momentum and the total spin
operator defined as $\bm{S}=\bm{S}_Q+\bm{S}_q$.
In a nonrelativistic constitutent-quark potential model, the
heavy-light quark potential consists of two different  
contributions: the confining linear potential
\begin{align}
  \label{eq:2}
V_{\mathrm{conf}}(r) = \varkappa\, r
\end{align}
with the parameter of the string tension $\varkappa$ and the
Coulomb-like interaction arising from one-gluon exchange
\begin{align}
  \label{eq:3}
  V_{\mathrm{Coul}}(r)=-\frac{4\alpha_{\mathrm{s}}}{3r},
\end{align}
where $\alpha_{\mathrm{s}}$ is the strong running coupling constant at
the one-loop level
\begin{align}
  \label{eq:4}
\alpha_{\mathrm{s}}(\mu) =  \frac{1}{\beta_0}
  \frac{1}{\ln(\mu^2/\Lambda_{\mathrm{QCD}}^2)}.
\end{align}
The one-loop $\beta$ function is given as
$\beta_0=(33-2N_f)/(12\pi)$. The dimensional transmutation parameter
are taken from the Particle Data Group (PDG)~\cite{PDG}, i.e. 
$\Lambda_{\mathrm{QCD}}=0.217\,\mathrm{GeV}$.
Since we include the charmed quark, the number of flavor is given by
$N_f=4$. The scale parameter $\mu$ will be set equal to the mass of
the charmed quark.
\begin{align}
  \label{eq:5}
V_c(r) = V_{\mathrm{conf}}(r) + V_{\mathrm{Coul}}(r)
\end{align}
and the spin-dependent parts are generated from this central
  potential and are expressed as
\begin{align}
V_{SS}(r)&=\frac{32\pi\alpha_{\mathrm{s}}}{9M_Q M_q}
           \delta(\bm{r}), \cr
V_{LS}(r) &= \frac{1}{2M_Q M_q} \left(\frac{4\alpha_{\mathrm{s}}}{r^3}
            - \frac{\varkappa}{r}\right), \cr
V_{T}(r)&=\frac{4\alpha_{\mathrm{s}}}{3 M_Q M_q}\frac{1}{r^3},
\label{eq:6}
\end{align}
where $M_Q$ and $M_q$ are stand for the dynamical heavy and light  
quark masses, respectively, which will be discussed shortly.  

In a practical calculation, the point-like spin-spin interaction is
required to be smeared by using the exponential form 
\begin{align}
\delta_{\sigma}(r)=(\frac{\sigma}{\sqrt{\pi}})^3e^{-\sigma^2r^2},
\label{eq:sigma}
\end{align}
where $\sigma$ stands for the smearing factor. 
Thus, one has a given set of parameters $\varkappa$ and $\sigma$ which
are fit to the spectra of mesons. In order to reduce the number of
free parameters in the present work, we fix the strong running
coupling constant $\alpha_s=0.4106$ defined in Eq.~\eqref{eq:4} at the
the scale of the charmed quark mass:
$\mu=M_Q=m_c^{\mathrm{current}}+\Delta M_Q$ with
$m_c^{\mathrm{current}}=1.275$\,GeV and $\Delta M_Q=0.086$\,GeV. Here
$\Delta M_Q$ is the shift of the heavy quark mass caused by the
heavy-light quark interactions that arise from a random instanton gas
of the QCD vacuum. Its numerical value used here is determined in
Ref.~\cite{Chernyshev:1995gj} (see also discussions in
Ref.~\cite{Turimov:2016adx}). The dynamical mass of the light quark
arises from the spontaneous breakdown of chiral symmetry
(SB$\chi$S). The QCD instanton vacuum explains quantitatively the
mechanism of the $SB\chi$S~\cite{Diakonov:1985eg} (see also
reviews~\cite{Schafer:1996wv,Diakonov:2002fq}). In the present work,
we take the value of $M_{\mathrm{u,\,d}}=340\,\mathrm{MeV}$. The
strange dynamical quark mass is taken to be
$M_{\mathrm{s}}=m_{\mathrm{s}}+M_q=(150+340)\,\mathrm{MeV}=\,490\,\mathrm{MeV}$.   

Since the main purpose of the present work is to consider
the contribution of the nonperturbative heavy-light quark interaction
from the instanton vacuum, we will introduce the effective
instanton-induced heavy-light quark potential. For simplicity, 
we follow Ref.~\cite{Chernyshev:1995gj}, where the local effective
interactions between the heavy and light quarks due to instantons were
derived in terms of the heavy and light quark operators $Q$ and $q$
\begin{align}
  \mathcal{L}_{qQ}=&-\left(\frac{M_q \Delta M_Q}{2n N_c}\right)
\left(\overline{Q} \frac{1+\gamma^0}{2} Q \overline{q}q
 +\frac14 \overline{Q} \frac{1+\gamma^0}{2}\lambda^a Q \overline{q}
  \lambda^a q\right),\cr
\mathcal{L}_{qQ}^{\mathrm{spin}} =&-\left(\frac{M_q \Delta
                                   M_Q^{\mathrm{spin}}}{2n N_c}\right)
 \frac14 \overline{Q} \frac{1+\gamma^0}{2} \lambda^a
 \sigma^{\mu\nu} Q \overline{q}    \lambda^a   \sigma_{\mu\nu} q.
\end{align}
The density parameter $n$ of the random instanton gas is defined by 
$N/2V_4 N_c$, where $N/V_4\sim 1\,\mathrm{fm}^{-4}$ is the instanton
density with the four-dimensional volume $V_4$ and $N_c$ denotes the 
number of colors. $\Delta M_Q$ is the mass shift of the heavy quark
caused by the instantons. $\Delta M_Q^{\mathrm{spin}}$ arises from the
$M_Q^{-1}$-order chromomagnetic interaction and, therefore, its
value is different from that of $\Delta M_Q$. 
In Ref.~\cite{Chernyshev:1995gj}, the numerical value of $\Delta
M_Q^{\mathrm{spin}}$ is determined to be $3\,\mathrm{MeV}$ for the
charmed quark.  Other standard quantities in the Lagrangian are the
Gell-Mann matrices for color space and the combinations from the
Dirac matrices. Consequently, the relevant two-body
instanton-induced central and spin-spin potentials are expressed as
\begin{align}
V^c_{\mathrm{I}}(\bm{r}) &=  \left(\frac{M_q\Delta M_Q}{2n
                    N_c}\right)
 \left( 1 +   \frac14 \lambda_q^a  \lambda_Q^a \right)
  \delta^3   (\bm{r}),
\label{eq:8}\\
V^{\mathrm{spin}}_{\mathrm{I}} (\bm{r}) &= -\left(\frac{M_q \Delta
               M_Q^{\mathrm{spin}}}{2n N_c}\right)
  \bm{S}_q \cdot \bm{S}_Q \lambda_q^a
  \lambda_Q^a  \delta^3 (\bm{r}),
\label{eq:9}
\end{align}
where $\bm{r}$ designates the relative coordinates $\bm{r}
=\bm{r}_q-\bm{r}_Q$.

Yet another spin-dependent potentials~\cite{Eichten:1980mw} are
derived from the central potential from the instanton vacuum as
follows:
\begin{align}
V_{SS}^I(r)&=\frac{1}{3M_QM_q}\nabla^2V_I(r), \cr
V_{LS}^I(r)&=\frac{1}{2M_QM_q}\frac{1}{r}\frac{dV_I(r)}{dr},\cr
V_T^I(r)&=\frac{1}{3M_QM_q}\left(\frac{1}{r}
          \frac{dV_I(r)}{dr}-\frac{d^2V_I(r)}{dr^2} \right),
\end{align}
Since the central and spin-spin potentials are given as the Dirac
delta functions, we need to introduce here also a 
smearing function to remove any divergence that would be caused by
them. So, we introduce the Gaussian type of the smearing function
\begin{align}
\delta_{\sigma_I}(r)=(\frac{\sigma_I}{\sqrt{\pi}})^3e^{-\sigma_I^2r^2}
\label{eq:sigmaI}
\end{align}
in both central and spin-spin potentials. Here $\sigma_I$ stands
for the another smearing factor, of which the numerical value will not
be much changed from that of $\sigma$ to avoid any additional
uncertainty. The explicit forms of the spin-dependent potentials are
obtained as 
\begin{align}
V_{SS}^I(r)=& \left(\frac{\Delta M_{Q}}{6nN_{c} M_Q}
              \right)
              \left(1+\frac{1}{4}\lambda_{Q}^{a}\lambda_{\bar{q}}^{a}
              \right)
\left(-6\sigma_I^2+4\sigma_I^4r^2
  \right)\delta_{\sigma_I}(r), \cr
V_{LS}^I(r)=& \left(\frac{\Delta M_{Q}}{4nN_{c} M_Q} \right)
              \left(1+\frac{1}{4}\lambda_{Q}^{a}\lambda_{\bar{q}}^{a}
              \right) (-2\sigma_I^2)\delta_{\sigma_I}(r),\cr
V_T^I(r)=& \left(\frac{\Delta M_{Q}}{6nN_{c} M_Q}
           \right)
           \left(1+\frac{1}{4}\lambda_{Q}^{a}\lambda_{\bar{q}}^{a}
           \right) (-4\sigma_I^4r^2)\delta_{\sigma_I}(r).
\label{Eq:12-14}
\end{align}

The total potential can be constructed by combining the potentials 
from the instanton vacuum given in Eqs.~\eqref{eq:8}, \eqref{eq:9},
and \eqref{Eq:12-14} with those from
the confining and Coulomb-like potentials in Eqs.~\eqref{eq:5} and
\eqref{eq:6}
\begin{align}
V_{Q\bar{q}}(r) = V(r) + V_I(r).
\end{align}
where $V_I(r)$ is defined as
\begin{align}
V_I(r) = V_{\mathrm{I}}^c (r)
+V_I^{\mathrm{spin}}(r)+V_{SS}^I(r)(\bm{S}_Q\cdot
\bm{S}_q)+V_{LS}^I(r)(\bm{L}\cdot\bm{S}) + V_{T}^I(r) [3(\bm{S}_Q\cdot
\hat{\bm{n}})(\bm{S}_q\cdot \hat{\bm{n}})- \bm{S}_1\cdot \bm{S}_2].
\end{align}
The matrix element of the potential in the ${}^{2S+1}L_J$ basis is
given by 
\begin{align}
\langle {}^{2S+1}L_J|V_{Q\bar{q}} (\bm{r}) |  {}^{2S+1}L_J\rangle
&=\tilde{V}_{c}(r)+\left[\frac{1}{2}S(S+1)-\frac{3}{4}\right]
  \tilde{V}_{SS}(r)
 +\frac{1}{2}\langle \bm{L}\cdot \bm{S}\rangle \tilde{V}_{LS}(r)  \cr 
& +\left[-\frac{2\langle \bm{L} \cdot \bm{S} \rangle (2\langle \bm{L}
  \cdot \bm{S}\rangle+1)}{4(2L-1)(2L+3)} 
 +\frac{S(S+1)L(L+1)}{3(2L-1)(2L+3)}\right] \tilde{V}_T(r),
\end{align}
where
\begin{align}
\langle \bm{L}\cdot \bm{S}\rangle=[J(J+1)-L(L+1)-S(S+1)]/2.
\end{align}
 Here we have taken the conventional spectroscopic notation
${}^{2S+1}L_J$ given in terms of the total spin $S$, the orbital
angular momentum $L$, and the total angular momentum $J$ with the
addition of the angular momenta, $\bm{J}=\bm{L}+\bm{S}$.
The corresponding terms $\tilde{V}_{c}(r)$, $\tilde{V}_{SS}(r)$,
$\tilde{V}_{LS}(r)$ and $\tilde{V}_T(r)$ denote generically the
central, spin-spin, spin-orbit, and tensor parts of the total
potential. 

\section{Calculations and Results}
In Ref.~\cite{Chernyshev:1995gj}, the mass spectra of the heavy
mesons were already studied within a simple variational method, the
potential from the instanton vacuum and the potential of the simple
harmonic oscillator being combined. The results from
Ref.~\cite{Chernyshev:1995gj} were in qualitative agreement with the 
experimental data. However, it is essential to consider more realistic 
contributions such as the confining potential and the Coulomb-like
potential from one-gluon exchange in order to understand the effects
of the instantons on the mass spectra of the heavy mesons in a
quantitative manner. In the present work, we will include all the
potentials mentioned in the previous section.

A nonrelativistic potential approach for a heavy-light quark system
is represented by the time-independent Schr\"odinger equation
with the static potential $V_{Q\bar q}(\bm{r})$
\begin{align}
\left[-\frac{\hbar^2}{\tilde{\mu}}\nabla^2+V_{Q\bar
  q}(\bm{r})-E\right]\Psi_{JM}(\bm{r})=0,
\label{eq:Schroe}
\end{align}
where $\tilde{\mu}$ denotes the reduced mass of the heavy meson system
and $\Psi_{JM}$ stands for the wavefunction of the state with
the total angular momentum $J$ and its third component $M$.
To solve the Schr\"{o}dinger equation numerically,
we employ the GEM which was successfully
applied to describe few-body systems such as light nuclei~(see a
review~\cite{Hiyama:2003cu} and references therein).

In the GEM the wavefunction is expanded in terms of a set of
$L^2$-integrable basis functions
$\{\Phi_{JM,k}^{LS};\,k=1-k_{\mathrm{max}}\}$ 
\begin{align}
\Psi_{JM}(\bm{r})=\sum_{k=1}^{k_{\mathrm{max}}}
  C_{k,LS}^{(J)}\Phi_{JM,k}^{LS}(\bm{r})
\end{align}
and the Rayleigh-Ritz variational method is used. So, we are able to
formulate a generalized eigenvalue problem given as
\begin{align}
\sum_{m=1}^{k_{\mathrm{max}}}
  \left\langle\Phi_{JM,k}^{LS}\left|-\frac{\hbar^2}{\tilde{\mu}}
  \nabla^2 +V_{Q\bar{q}}(\bm{r})-E\right|\Phi_{JM,m}^{LS}\right\rangle
  C_{m,LS}^{(J)}=0\,.
\end{align}
The angular part of the basis function $\Phi_{JM,k}^{LS}$ is
expressed in terms of standard spherical harmonics and the normalized
radial part $\phi_{k}^{L}(r)$ is written in terms of the Gaussian
basis functions 
\begin{align}
\phi_{k}^{L}(r) = \left(\frac{2^{2L+\frac{7}{2}}
  r_{k}^{-2L-3}}{\sqrt{\pi}(2L+1)!!}\right)^{1/2}
r^{L}e^{-(r/r_{k})^2},
\end{align}
where $r_k,\, k=1,2,...,k_{max}$ designate variational parameters.
When it comes to the case of a two-body problem, the total number of
the variational parameters is reduced by using the geometric
progression in the form of $r_k=r_1a^{k-1}$, which provides a good
convergence of the results. Thus, in the two-body problem, we need
only three variational parameters, i.e. $r_1$, $a$ and
$k_{\mathrm{max}}$.\footnote{For more details,
see Refs.~\cite{Kamimura1988, Hiyama:2003cu, Hiyama2004, Hiyama2009,
  Hiyama-hyper}.}
Once the Schr\"{o}dinger equation is solved, the energy eigenvalue
$E_N$ is found and the mass of the heavy meson is determined by
\begin{align}
M= M_Q + M_q + E_N +\Delta E_{q},
\end{align}
where $\Delta E_{q}$ is the overall energy shift in the spectra
depending on the light-quark content of the meson and plays a role of
a simple tuning parameter. As mentioned already, $M_Q$ and $M_q$ are
the dynamical masses of the heavy and light quarks, respectively. Note
that $M_Q$ contains also the mass shift arising from the instanton
vacuum. In this work we will slightly vary the total mass of the
strange quark mass $M_s$ and try to analyze the corresponding
effects. 

Since, some of remaining parameters cannot be determined
theoretically, we construct several sets of the parameters and
call them Model I$^{\prime}$, Model I, Model II, and Model III,
respectively. The numerical values of model parameters are listed in
Table~\ref{tab:1} and we use them to calculate the spectra
of the heavy mesons.\footnote{The corresponding explanation of model 
parameters will be given hereafter in the text.}

\begin{table}[htp]
\caption{
Free parameters of the model: $m_s$ denote the
  dynamical mass of the strange quark, 
  $\varkappa$ stands for the string tension, $\sigma$ and  $\sigma_I$
  designate the smearing   parameters corresponding to point like
  interactions in   Eqs.\,(\ref{eq:sigma}) and\,(\ref{eq:sigmaI}), 
   $\Delta E_{u,d}$ and $\Delta E_s$ are the constant overall
  energy shifts of mesons corresponding to the up (down) and
  strange quark constituents, and $n$ is the density of instanton
  medium.} 
\begin{tabular}{cccccccc}
  \hline\hline
 Model & $m_{s}\,[\mathrm{GeV}]$ &
 $\varkappa\,[\mathrm{GeV}^2]$ &
$\sigma\,[\mathrm{GeV}]$ & $\sigma_I\,[\mathrm{GeV}]$ & $\Delta
E_{u,d}\,[\mathrm{GeV}]$  &$\Delta E_{s}\,[\mathrm{GeV}] $&
 $n\,[\mathrm{fm}^{-4}]$\\
  \hline
 I$^\prime$&0.450
&0.169&1.43&$-$&-0.365&-0.299&$-$
\\
 I&0.450
&0.169&1.43&1.18&-0.365&-0.299&1.0\\
 II & 0.490
&0.165&0.95&1.19&-0.347&-0.287&1.0\\
III & 0.470
&0.163&0.93&1.17&-0.339&-0.274&0.9\\
  \hline\hline
\end{tabular}

\label{tab:1}
\end{table}

The results of the charmed meson masses corresponding to
the different models are listed in Table~\ref{tab:2} in comparison
with the experimental data taken from the Particle Data Group
(PDG)~\cite{PDG}. In the second column, the results without 
instanton-induced quark-quark interactions
are presented. It is called Model $\mathrm{I}'$ that is obtained by
including only the confining and Coulomb-like type interactions. One
can assume that in this model the nonperturbative effects are only
taken into account by means of dynamically
generated masses of the corresponding light quarks.
It is seen that the results are relatively in good agreement with 
the experimental data. It indicates that a nonrelativistic approach to
the heavy-light quark system works even quantitatively at least for
the mass spectra of the conventional heavy mesons. 

Model I has the same parameter set as Model $\mathrm{I}'$ except for
the instanton-induced potentials, which means that the parameters are
not tuned but the instanton-induced heavy-light quark interactions are
taken into account. By doing this, we can examine how the 
instanton-induced quark-quark interactions affect the mass of each
charmed meson. The effects of instanton-induced
quark-quark interactions are clearly seen in the ground state
$D^\pm$ meson, while they are rather tiny on other charmed
mesons. In particular, the effects are almost negligible on the
$P$-wave charmed meson spectra. One can conclude that in general
instanton-induced interactions do not affect much the spectra
of heavy mesons and play only a role in the fine-tuning level.

Thus, we present the results of Model II in which the free parameters
are fitted to the experimental data. One can see that the results
slightly change in comparison with the Model I$^\prime$
and shows that the instanton-induced quark-quark
interactions are seem to be important in the fine-tuning level.
In Model III, we change also the density of the instanton medium is
slightly changed, considering it as an input parameter. This is
allowed, as was already discussed in Ref.~\cite{Turimov:2016adx} in
detail. All other parameters are fitted to the experimental data
as in the case of Model~II.

\begin{table}[htp]
\caption{The results of the charmed $D$-meson masses in units of
  MeV. The second column lists the results without the instanton-induced
quark-quark interactions and is coined as Model $\mathrm{I}'$. The
third, fourth, and fifth columns list those of Models I, II, and
III. The last column shows the corresponding experimental data taken
from PDG~\cite{PDG}.} 
\begin{tabular}{cccccc}
  \hline\hline
  Model& $\mathrm{I}'$ & I & II & III & Exp.\\
  \hline
  $D^\pm(1^1S_0)$ &1867.7&1787.0&1868.3& 1868.0& $1869.65\pm 0.05$ \\
  $D^{*\pm}(2^1S_0)$&2013.5&2006.4&2009.7&2010.2& $2010.26\pm 0.05$\\
  $D_1(1^1P_1)$ & 2461.2&2461.5&2458.7&2456.7& $2423.2\pm 2.4$  \\
  $D_2^*(1^3P_2)$ &2462.2&2461.2&2461.7 &2460.1& $2465.4 \pm 1.3$ \\
  $D^*(1^3S_1)$ & 2639.0&2593.4&2634.1&2630.4 &$2637 \pm 2 \pm 6$ \\
  $(2^3S_1)$ &2737.0&2732.6&2724.0&2719.8&  \\
  \hline \hline
\end{tabular}
\label{tab:2}
\end{table}
The results of Model III are slightly better than
those of Model II. As expected from the comparison of Model I with
Model $\mathrm{I}'$, the prediction of Model III is not much different
from that of Model I$^\prime$. Thus the potential from the instanton
 vacuum in the present form change slightly the mass spectrum
of the charmed mesons and does not affect quantitatively the
results from the calculation without instanton-induced
quark-quark  interactions. 

\begin{table}[htp]
\caption{The results of the charmed strange $D_s$-meson
masses in units of  MeV.
Other notations are same as in the case of Table~\ref{tab:1}.}
\begin{tabular}{cccccc}
  \hline \hline
  Model& $\mathrm{I}'$ & I & II & III& Exp.\\
  \hline
$D_s^\pm(1^1S_0)$ &1969.1&1887.9&1969.0 &1968.9& $1968.34 \pm 0.07$ \\
$D_s^{*\pm}(2^1S_0)$&2108.3 &2100.8&2113.2  &2111.5& $2112.2 \pm 0.4$\\
$D_{s1}^\pm (1^1P_1)$ &2538.3&2538.1&2543.1 &2540.5& $2535.10\pm0.06$\\
$D_{s2}^*(1^3P_2)$&2546.2&2545.1&2555.2&2551.8& $2569.1\pm 0.8$ \\
$D_s^*(1^3S_1)$ &2703.7&2661.6&2697.4 &2696.0& $2708.3_{-3.4}^{+4.0}$\\
$(2^3S_1)$&2792.6&2788.2&2780.5 &2778.5 &  \\
  \hline \hline
\end{tabular}
\label{tab:3}
\end{table}
Table~\ref{tab:3} lists the results of the charmed strange meson
masses. As done in Table~\ref{tab:2}, we first compute the masses of
the charmed strange mesons without the instanton contributions, which
are listed in the second column of Table~\ref{tab:2}. Then we
include the instanton-induced quark-quark interactions, of which the
results are presented in the other columns. 
The effects of the instantons are similar to the case of the
charmed mesons, that is, the instanton effects are noticeable only on
the ground state $D_s^\pm$ meson whereas they are negligibly 
small on the $P$-wave charmed strange mesons. Though the results of
Model III seem slightly better than those of Model I$^\prime$, for the
quark-quark potential from the instanton vacuum,
at least in the present form, the improvement is marginal
in the charmed strange meson mass spectrum. Moreover, the effects of
the instanton-induced potential on the charmed strange mesons are even
smaller than on the charmed nonstrange ones.   

Finally, we would like to note that although we have changed
the density of instanton medium $n$ in Model~III in comparison with
Model~II the mass contribution $\Delta M_Q$ is unchanged
and kept in both cases equal to 0.086\,GeV. However, 
$\Delta M_Q$ is proportional to $n$ and therefore it must
be also modified if the value of $n$ changes. As a result,
eigenfunctions and eigenvalues of the Hamiltonian should be also
altered. Consequently, a better fine-fitting of the whole mass 
spectra can be achieved by means of changes of instanton parameters in
a self-consistent manner.  Though these selfconsitent changes 
of parameters are expected to improve the present results further,
we do not perform it because in the present work we aim at examining
the effects of the existing nonperturbative heavy-light quark
potentials from the instanton vacuum on the conventional heavy
mesons. 
 
\begin{table}[htp]
\caption{The results of the instaton effects on the low-lying charmed
  heavy mesons in units of MeV. The values of the relevant parameters
  are taken from those for Model I.}
\begin{tabular}{cccccc}
  \hline\hline
Heavy meson  & Instanton contribution [MeV] & Exp. [MeV]\\
  \hline
  $D^\pm(1^1S_0)$  & 80.7& $1869.65\pm 0.05$ \\
  $D^{*\pm}(1^3S_1)$ & 7.1& $2010.26\pm 0.05$\\
  $D_1(1^1P_1)$  & -0.3& $2423.2\pm 2.4$  \\
  $D_2^*(1^3P_2)$ & 0.1& $2465.4 \pm 1.3$ \\
  $D^*(2^1S_0)$   &45.6&$2637 \pm 2 \pm 6$ \\
  $(2^3S_1)$  & 4.4&  \\
$D_s^\pm(1^1S_0)$ & 81.2& $1968.34 \pm 0.07$ \\
$D_s^{*\pm}(1^3S_1)$ & 7.5& $2112.2 \pm 0.4$\\
$D_{s1}^\pm (1^1P_1)$ & 0.2& $2535.10\pm0.06$\\
$D_{s2}^*(1^3P_2)$ & 1.1& $2569.1\pm 0.8$ \\
$D_s^*(2^1S_0)$ &42.1& $2708.3_{-3.4}^{+4.0}$\\
$(2^3S_1)$ &4.4 &  \\
  \hline \hline
\end{tabular}
\label{tab:4}
\end{table}
In Table~\ref{tab:4}, we list the results of the contributions from
the instanton-induced potentials. While they have visible effects on
the masses of the $D^{\pm}$ and $D_s^{\pm}$ mesons, and marginal
contributions to the radially excited $S$-wave $D^*(2^1 S_0)$ and
$D_s^*(2^1 S_0)$ mesons, they have almost no impact on other excited
$D$ and $D_s$ mesons. Thus, in conclusion, the present form of the
instanton-induced potentials contributes to some of the $D$ and $D_s$
mesons as explicitly shown in Tables~\ref{tab:2}, \ref{tab:3}, and
\ref{tab:4}, its overall effects turn out to be marginal.  
Possible ways of improving the present results will be mentioned in
the next Section.

\section{Summary and outlook}
In the present work, we have investigate the effects of the
heavy-light quark potential from the instanton vacuum on the mass
spectra of the conventional charmed mesons. First, we have considered
the confining potential that is proportional to the relative distance
between the heavy and light quarks. The Coulomb-like potential, which
arises from one-gluon exchange, has been included. The spin-dependent
potentials were generated from the central part. Then we have computed
the mass spectra of the charmed mesons, employing the Gaussian
expansion method to solve the nonrelativistic Schr\"{o}dinger
equation. The results are in good agreement with the experimental data
even without the potential from the instanton vacuum included. Then,
we have introduced the central and spin-dependent potentials from the
instanton vacuum. The additional spin part of the potential was
obtained from the central part of the instanton-induced
potential. While the instanton effects are noticeable on the $S$-wave
charmed and charmed strange heavy mesons, the contribution from the
instanton-induced potential is rather tiny to their masses.

Though the present form of the instanton-induced potential does not
give any significant contribution to the heavy meson masses, there are
some possible ways of elaborating the present analysis:
\begin{itemize}
\item The present work is based on the nonrelativistic
  Schr\"{o}dinger equation, since we aim mainly at investigating the
  effects of the instanton-induced potential. However, once the light
  quark is involved, it is inevitable to include certain relativistic
  effects.
\item The instanton-induced potentials used in the present work was
  derived from the random instanton gas model and are given as local
  ones. However, if one uses the instanton liquid model, the
  interaction between the heavy and light quarks turn out to be
  nonlocal~\cite{Musakhanov:2017gym}. This nonlocality will have
  certain effects on the mass spectra of the heavy mesons.
\item Recently, Ref.~\cite{Musakhanov:2017erp} showed that
  rescattering of gluons with instantons generates dynamically the
  effective momentum-dependent gluon mass that will cause the screened
  heavy-quark potential. It indicates that certain nonperturbative
  effects from the instanton vacuum will contribute also to the
  heavy-light quark system.
\end{itemize}
Thus, one needs to study systematically nonperturbative effects
on both heavy mesons and heavy baryons, arising from the instanton
vacuum. The corresponding investigations are under way.

\acknowledgments
H.-Ch.K is grateful to P. Gubler, A. Hosaka, T. Maruyama and M. Oka
for useful discussions.  He wants to express his gratitude to the
members of the Advanced Science Research Center at Japan Atomic Energy
Agency for the hospitality, where part of the present work was done. 
The work of QW is supported by National Natural Science Foundation
of China (11475085, 11535005,11690030) and National Major state Basic
Research and Development of China (2016YFE0129300).
The works HChK and UY are supported by Basic Science Research Program
through the National Research Foundation (NRF) 
of Korea funded by the Korean government (Ministry
of Education, Science and Technology, MEST), Grant
No. NRF2018R1A2B2001752 (HChK) and Grant No. 2016R1D1A1B03935053 (UY).

\end{document}